Commentaries on the publication entitled

Evidence for continuity of interstitial spaces across tissue and organ boundaries in humans

Hongyi Li [1]

[1] Beijing Hospital, National Center of Gerontology, Beijing, China

**Corresponding author:**

Hongyi Li, Cardiology Department, Research Center for Interstitial Fluid Circulation & Degenerative Diseases and Aging, Beijing Hospital, 100730 Beijing, China. Email: leehongyi@bjhmoh.cn;

Interstitial connective tissues are one of the four basic types of animal tissue and continuously distributed throughout the body. Illustrated by the movements of the 0.41-0.64μm particles across the defined tissue boundaries and the continuously distributions of the fluorescently stained hyaluronic acid within the fibrous tissues around blood vessels and nerves, the authors of this paper suggested that interstitial fluid (ISF) is not fixed locally but would flow through fibrous matrices for a long-distance, even between more distant parts of the body.

Since Ernest Starling described the net flow of fluid between the connective tissue spaces and the capillary lumen in 1896, the transport pattern and mechanical mechanism of ISF flow through extravascular spaces or interstitial connective tissues have been explored by generations. The components of interstitial connective tissues are revealed to be the gel-like complex of the fibrous network, glycosaminoglycans (GAG), proteoglycan and interstitial spaces for fluid flow, which is highly resistant to hydraulic flow from capillary to lymphatic vessels[1]. The measurement of the hydraulic conductivity of interstitial matrices requires determination of the flow versus pressure relation across a layer of tissues of known dimensions[1]. Based on the methods designed by Guyton and Levick, it was found that the hydraulic conductivity is small and correlates negatively with the GAG and collagen concentration over a wide variety of tissues[1]. Therefore, the ISF diffuses mainly for short distances through the gel-like matrix between capillaries and adjacent cells. Nevertheless, when a dye is injected into the circulating blood, it often can be seen to flow through the interstitium in the small rivulets, usually coursing along the surfaces of collagen fibers or surfaces of cells[2]. In conventional physiology, it is believed that the amount of "free" fluid is slight and usually less than 1 percent in normal tissues[2]. Intriguingly, whereabout in interstitial connective tissues does the "less free fluid" flow? By comparison with the intricate measurements of interstitial pressure, it is an alternative to use an imaging tracer for direct observations of the ISF flow.

In the 1950s-1970s, a linear fluorescent pathway was visible along the elastic fibers of the interstitial connective tissues after the passage of fluorescent dye through the microvessel wall in the mesentery of the rabbit and cat[3]. The fluorescein transport occurs much faster than diffusion and the elastic fibers might have a passive transport or guide rail function for fluid flow between arterial and venous regions of the capillaries and lymphatic vessels[3]. The fluorescein transport pathway in interstitial connective tissues was marked as low resistance pathway or described as an extravascular fluid pathway by Kihara in 1956[3].

By means of incomplete dark field transilluminations and electron microscopy, the transinterstitial movements of the marked fluid were named as prelymphatic or interstitial tissue channels by Catchpole, Casley-Smith and G. Hauck in the 1970s, respectively[3]. The tissue channels are considered to connect continuously with the initial lymphatic vessels, forming a random network of converging drainage pathways in interstitial connective tissues[4]. Illustrated by India ink suspension, fluorescein-isothiocyanate, ferrocyanide precipitates, the positions of the tracers are found to be near vessel wall or in the skeletal muscle tissues or intestinal wall and represent a water-rich region in interstitial connective tissues[3,4,5]. The tissue channels range from 30nm to 50μm in diameter and play a role in controlling small molecules to penetrate interstitial connective tissues[3,4,5]. However, the spatial structures of such a channel and its surroundings in the gel-like interstitial matrix for fluid flow have not been adequately stained and identified until now. Due to the defined deficiencies, the understanding on tissue channel is not clearly enough to distinguish a continuous interstitial space in the meshwork of fibrous connective tissues for a long-distance ISF flow like the systemic flow of blood and lymph.

In the 1800s, the perivascular spaces (PVS) were described by Durand-Fardel, Rudolf Virchow and Charles Robin respectively[6]. The PVS include a variety of passageways or small, linear, fluid-filled structures around arterioles, capillaries and venules in the brain, along which fluid and substances can move[6]. The boundary structures of PVS are meningeal membranes and seem to be an enlarged space along the outside of a vascular vessel rather than a fibrous mesh of the connective tissues[6]. The driving mechanism of fluid flow in the PVS seems to depend on vascular pulsation, respiratory movement, the sleep-wake cycle and AQP4 water channels[6].

This publication claimed there are a structural continuity of fibrous tissues covering on both nerves and blood vessels, including the perineurium, vascular adventitia and layers of the adjacent fascia on them, in which a continuous body-wide interstitial space is formed for the ISF flow. Nevertheless, whether their findings are the results of the ISF flow in a PVS-like pathway or throughout the adventitial tissues along the vessels and nerves needs further explorations.

In 2012 by MRI and real-time fluorescence stereomicroscopy in alive rabbits, the ISF flow along venous adventitia was found along the lower extremity veins, inferior vena cava of the abdomen and thorax, and into three grooves of the heart, contributing to form pericardial fluid[7]. Such an adventitial ISF flow includes two locations: fluid flow through the tunica adventitia and through the perivascular connective tissues along the outside of the vessels[7]. In an *ex vivo* human lower leg sample, the anatomical structures of a long-distance ISF flow pathway from ankle dermis were identified to include at least 4 types by fluorescence imaging methods: the cutaneous-, perivenous-, periarterial-, and neural- ISF flow pathways, each of which is conformed to be the fibrous connective tissues[8]. In human cadavers, the structural framework of the long-distance perivenous ISF flow pathways was found to be the highly ordered and topologically connected solid fibers in interstitial matrix, acting as a guiderail for the long-distance ISF flow[9]. The experimental findings on animals, the amputated lower legs and human cadavers have disclosed a unique kinetic and dynamic pattern of a long-distance ISF flow: the ISF can be "pulled" from the long-distance fibrous pathways into a driving center, such as the beating heart[10].

In summary, increasing data indicate the ISF might not only diffuse locally but also circulate in the network of interstitial connective tissues systemically around the whole body. The fluid-filled interstitial spaces are similar with the interstitial tissue channel, lacking a distinct spatial structure for "free" interstitial fluid flow, which might be the key to comprehend the whereabouts and directions of a systemic ISF flow and circulation.